\documentclass[doublecol]{epl2}

\usepackage{amsmath}
\usepackage{amsthm}
\usepackage{amssymb}
\usepackage{graphicx}
\usepackage{subfigure}
\usepackage{cancel}
\usepackage{xr}
\graphicspath{{./figures/}}

\newcommand{\abs}[1]{\left| #1 \right|}
\newcommand{\mean}[1]{\left\langle #1\right\rangle}
\newcommand{\Om}{\Omega}
\newcommand{\Pe}{\mathrm{Pe}}

\title{Noisy pursuit by a self-steering active particle in confinement}

\author{Marielle Ga{\ss}ner, Segun Goh, Gerhard Gompper, and Roland G. Winkler}

\institute{                    
  Theoretical Physics of Living Matter, Institute of Biological Information Processing  and Institute for Advanced Simulation, Forschungszentrum J\"ulich and JARA, 52425 J\"ulich, Germany
}

\abstract{The properties of a cognitive, self-propelled, and self-steering particle in the presence of a stationary target are analyzed theoretically and by simulations. In particular, the effects of confinement in competition with activity and steering are addressed. The pursuer is described as an intelligent active Ornstein-Uhlenbeck particle (iAOUP), confined in a harmonic potential. For the free pursuer, we find universal scaling regimes for the pursuer-target distance in terms of the P\'eclet number and maneuverability. Steering results in a novel constant mean-distance regime, which broadens with increasing maneuverability. Confinement strongly affects the propulsion direction and leads to a scaling at large P\'eclet numbers similar to that in absence of confinement, yet with a pronounced dependence on confinement strength.
}

\begin{document}

\maketitle

\section{Introduction}

Sensing of the environment and adaptation of motion is vital for the survival of biological entities, ranging from microbes and cells on the microscale to animals on the macroscale \cite{vics:12,jike:15,bech:16,elge:15,lan:16,scha:18,seng:21,goh:22,couz:05}, and is fundamental for task-oriented motion of robots  \cite{pala:18,sitt:17,kasp:21,huan:22}. Evolution has provided a diverse spectrum of propulsion as well as sensing mechanisms, which differ fundamentally over the wide range of length scales. Yet, the same basic characteristics applies on all scales, which is the combination and interplay of sensing, response, and adaptation of motion. The adopted response allows for goal-oriented motion, which encompasses, for example, cell movement in wound healing, sperm navigation toward the egg \cite{jike:15}, and foraging and prey-searching activities of animals. Inspired by biology, various synthetic propulsion mechanisms have been designed, either exploiting biomimetic approaches \cite{pala:18} or applying novel strategies based on chemotaxis and thermotaxis \cite{bech:16}. Moreover, various sensing and steering strategies for colloidal systems have been implemented \cite{baeu:18,lave:19,selm:18,zhan:21,alva:21}.

Theoretical insight into the emergent properties of systems of self-propelled particles is provided by two generic models, the Active Brownian Particle (ABP) \cite{bech:16} and the Active Ornstein-Uhlenbeck Particle (AOUP) \cite{fodo:16,das:18.1}. Both models take into account self-propulsion, thermal or active noise, and, if required, conservative interactions. The major difference is their propulsion velocity, which is inherently constant for an ABP, whereas it is determined by a stochastic Ornstein-Uhlenbeck process \cite{risk:89} for an AOUP. Perception can be implemented via the cognitive flocking model \cite{barb:16,bast:20,goh:22}. It  assumes that the moving entities navigate by using exclusively the instantaneous ``visual information'', which they receive about the positions of other entities. Such an orientational adaptive force has been utilized in combination with ABPs, and leads to  steering  with limited maneuverability toward the direction of a sensed object \cite{barb:16,bast:20,goh:22,negi:22}. 

An optimal steering dynamics is often hampered by environmental factors such as geometrical obstacles, flow fields, or other forces \cite{zano:21,wyso:22}. Landscapes of external forces/fields can obviously be very complex and, hence, strongly affect the dynamics of the migrating active entity. 

In this article, we study -- both analytically and by computer simulations -- the pursuit dynamics of an active self-steering agent toward a stationary target in absence and presence of a confining potential. The pursuer is modeled as an AOUP  augmented by active reorientation of its velocity toward the target under the influence of noise (henceforth denoted as ``intelligent Active Ornstein-Uhlenbeck particle'', iAOUP).  The simple AOUP allows for an analytical solution of the equations of motion, even in the presence of a harmonic potential \cite{das:18.1}. Moreover, it yields the same second moments as the ABP model \cite{das:18.1,eise:17,cara:22}. For the iAOUP, steering is treated similar to the cognitive flocking model, however, adapted to the absence of the constraint of a constant propulsion velocity as for ABPs. We find that the iAOUP dynamics exhibits significant differences to that of an intelligent Active Brownian Particle (iABP) \cite{goh:22}. In particular, the scaling relations of the mean distance from the target  and the orientation of the propulsion direction toward the target in terms of the P\'eclet number $\mathrm{Pe}$ and maneuverability $\Om$ differ significantly from those of an iABP.  Even more, a new regime emerges, where the mean distance is independent of $\mathrm{Pe}$ and $\Om$. The adaptation of the additional degree of freedom -- the magnitude of the velocity in terms of a stochastic process -- and its coupling to the radial distance and orientation of the propulsion direction, can enhance the pursuit performance. As a paradigmatic case of the influence of a potential landscape on pursuit, we consider the active dynamics in the presence of a confining harmonic potential.  We find that confinement strongly affects the orientation of the propulsion direction and consequently the pursuer distance from the target.

\section{Model}

The two-dimensional dynamics of an iAOUP in the vicinity of the target located at the origin of the reference frame is described  by the overdamped Langevin equations 
\begin{subequations}
\label{eq:langevin_vec}
\begin{align}
	\dot{\vect{r}} &= \vect{v}  - \frac{1}{\gamma} \nabla U(\vect{r}) + \sqrt{2 D_T} \, \vect{\Gamma} \, ,\label{eq:langevin_vecr}\\
	\dot{\vect{v}} &= -D_R \vect{v} - C \frac{\vect{r}}{\abs{\vect{r}}} + v_0 \sqrt{D_R} \, \vect{\eta} \, ,\label{eq:langevin_vecv}
\end{align}
\end{subequations}
where $\bm r(t)$ is the pursuer position, $\bm v(t)$ the propulsion velocity, $\gamma$ the friction coefficient,  and 
$\bm \Gamma$ (t) and $\bm \eta(t)$ are Gaussian and Markovian stochastic processes of zero mean and variances $\left\langle \Gamma_{\alpha}(t) \Gamma_{\beta}(t') \right\rangle = \left\langle \eta_{\alpha}(t) \eta_{\beta}(t') \right\rangle = \delta_{\alpha \beta}\delta(t-t')$, with $\alpha, \beta \in \{x,y\}$.  $D_T=k_BT/\gamma$ and $D_R$ are the translational and rotational diffusion coefficients ($T$ is the temperature and $k_B$ Boltzmann's constant), and  $v_0$ is proportional to the average propulsion velocity of an AOUP. Confinement is described here by the radial harmonic potential  $U(\bm r) = \kappa r^2/2$. 

\begin{figure}[t]
\includegraphics[width =\columnwidth]{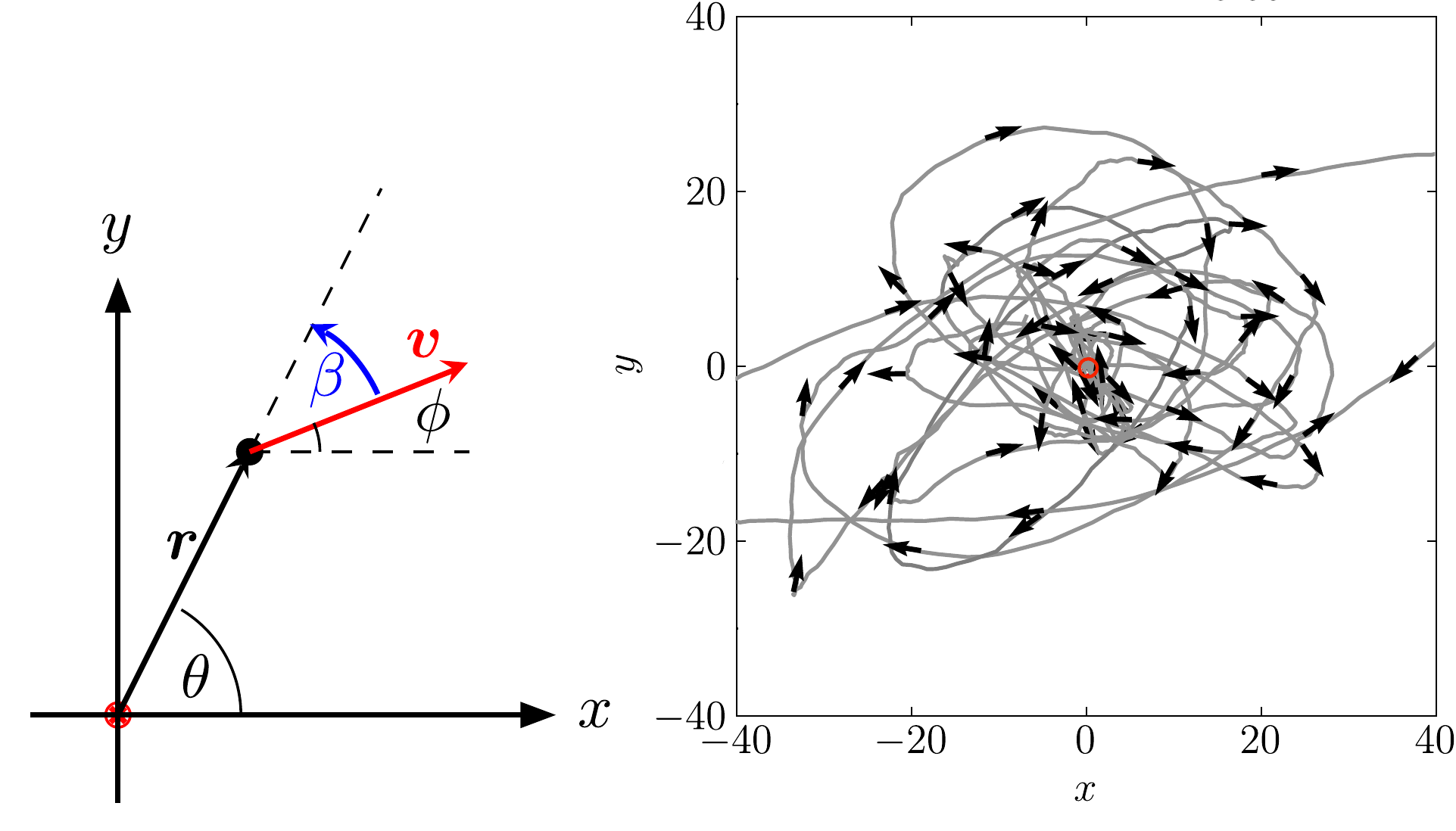}
\caption{(a) Geometry for the stationary target (red circle at origin) and  the pursuer (black bullet) at position $\bm r$ and with self-propulsion velocity $\bm v$. The (bearing angle) $\beta$ is defined as $\beta = \theta-\phi$. (b) Trajectory of the pursuer for $\mathrm{Pe} = 128$ and $\Om = 8$. The red circle in the center represents a region of radius $r_H$  and the  arrows point in the direction of the instantaneous propulsion velocity.}
\label{fig:sketch_geom}
\end{figure}

For a free iAOUP,  $\mean{\bm v^2}$ is equal to the mean-square propulsion velocity of an ABP, i.e., $\mean{\bm v^2} = \bm v_0^2$.  The active adaptation force $ -C \bm r/\abs{\vect{r}}$ orients the iAOUP propulsion velocity toward the target and also affects its magnitude -- accelerating it in the direction of the target. With the introduction of  polar coordinates $(r, \theta)$, $(v,\phi)$, and the angle $\beta = \theta - \phi$ -- denoted as bearing angle in the following --  between the position vector and propulsion velocity vector of the iAOUP (Fig.~\ref{fig:sketch_geom}(a)), Eqs.~\eqref{eq:langevin_vec} become (in the Stratonovich sense \cite{risk:89}) (see supplementary material)
\begin{subequations}
\label{eq:langevin} 
\begin{align}
	\dot{r} &= \mathrm{Pe} \, v \, \cos \beta - k r + \sqrt{2} \vect{e}_r \cdot \vect{\Gamma} \label{eq:langevin_r} \, ,\\
	\dot{v} &= - v - \Omega \cos \beta   + \vect{e}_v \cdot \vect{\eta} \label{eq:langevin_v} \, ,\\
	\dot{\beta} &= \left( \frac{\Om}{v} - \frac{v \, \mathrm{Pe}}{r} \right) \sin \beta + \left( \frac{\sqrt{2}}{r}  \vect{e}_{\theta} \cdot \vect{\Gamma}- \frac{1}{v} \vect{e}_{\phi} \cdot \vect{\eta} \right)  . \label{eq:langevin_beta}	
\end{align}
\end{subequations}
Here, we measure lengths, velocities, and time in units of $r_H = \sqrt{D_T/D_R}$, $v_0$, and $D_R$, respectively. The $\bm e_i$ represent the different polar unit vectors. Activity is expressed by the P\'eclet number $\mathrm{Pe} = v_0/ (r_H  D_R)$, the maneuverability by $\Om = C/(v_0 \,D_R)$, and the strength of the harmonic potential by $k= \kappa/(\gamma D_R)$. An example of a trajectory is presented in Fig.~\ref{fig:sketch_geom}(b). 

\section{Results: Pursuit by a free iAOUP}

\begin{figure*}[t]
\includegraphics[width = \textwidth]{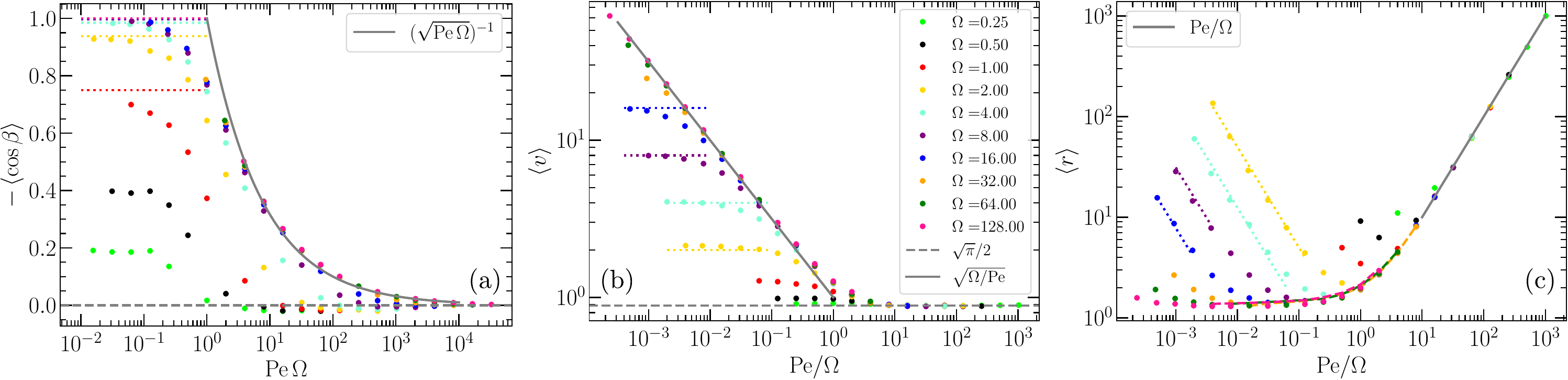}
\caption{Stationary-state properties of free  iAOUPs. 
(a) Mean bearing angle  $\mean{\cos\beta}$ as a function of $\mathrm{Pe}$ and various $\Om$. For $\mathrm{Pe} \, \Om \gtrsim 1$ and $\Om \gg1$, the simulation data (bullets) are described by Eq.~\eqref{eq:cos_universal} (gray line). The dotted lines represent Eq.~\eqref{eq:mean_cos_pe_small}.  
(b) Mean propulsion velocity $\mean{v}$ as a function of $\mathrm{Pe}$. For $\mathrm{Pe} \gg \Om$, the AOUP-limit $\mean{v} = \sqrt{\pi}/2$ is approached (gray dashed line).  For $\mathrm{Pe} \to 0$, $\mean{v} \approx \Om$, as indicated by the dotted lines. The power-law regime of Eq.~\eqref{eq:cos_universal} (gray line) is assumed for $\Om \gg 1$ and $\mathrm{Pe}/\Om \ll 1$.
(c) Mean radial distance $\mean{r}$ from the target as a function of $\mathrm{Pe}$. For $\mathrm{Pe}/\Om \gg 1$, the asymptotic dependence of Eq.~\eqref{eq:mean_r_high} (gray line) is assumed. The dashed lines represent  Eq.~\eqref{eq:mean_r_high}  with the velocity averages from simulations  and the dotted lines depict Eq.~\eqref{eq:mean_r_lowPe}. Characteristic  trajectories are presented in Fig.~1 and S8. (See also movies supplM1, supplM2.)}
\label{fig:mean_distance}
\end{figure*}

Figure~\ref{fig:mean_distance} displays simulation results for the dependence of $\langle \cos \beta \rangle$, the mean velocity $\mean{v}$, and the mean radial distance $\mean{r}$ on the P\'eclet number for various values of the maneuverability. Over certain regimes, the averages exhibit universal scaling behavior as a function of $\Pe \, \Om$ or $\Pe/\Om$, where $\Pe \, \Om = C/(r_H D_R^2)$ is independent of the  propulsion velocity and $\Pe/\Om = v_0^2/(r_H C)$ is independent of $D_R$.    
Theoretical insight into the iAOUP dynamics and approximations for the averages are obtained by the analysis of the nonlinearly coupled Fokker-Planck equations corresponding to Eqs.~\eqref{eq:langevin}, as presented in the Supplementary Material (SM).

Three regimes can be distinguished in Fig.~\ref{fig:mean_distance}, depending on the P\'eclet number, the maneuverability, and their ratio.
 
\subsection{Small P\'eclet number} In the limit $\mathrm{Pe} \to 0$ and $\Om \to 0$, the dynamics of a simple AOUP is obtained, which is determined by  thermal and active fluctuations, with $\mean{\cos\beta} =0$, $\mean{v}=\sqrt{\pi}/2$, and a constant radial distribution function $\Psi(r)$. The presence of steering leads to a competition with active propulsion.

With increasing maneuverability, but $\Pe \, \Om \ll 1$,  $|\mean{\cos \beta}|$ assumes a $\Pe$-independent plateau, which approaches the limiting value $\mean{\cos \beta} =-1$ for $\Om \gg 1$,  i.e., the propulsion direction is oriented toward the target (Fig.~S8 shows trajectories).  Our analytical calculations yield the average value  (SM, Sec.~S-III.B.)
\begin{equation}
\label{eq:mean_cos_low}
    |\mean{\cos \beta }| = \frac{I_1(z)}{I_0(z)} \, ,
\end{equation}
with  $z = 2 \mean{v}  \Omega$, and $I_0$, $I_1$ the modified Bessel functions of the first kind.  
Similarly, for $\Om |\mean{\cos \beta}| \gg 1$, the mean velocity approaches a $\Pe$-independent plateau, which is given by  
\begin{align} \label{eq:mean_v_small}
\mean{v} = \Om |\mean{\cos\beta}| \, , 
\end{align}
and follows from the general expression (S10) of Sec.~S-III.A. (SM, Fig~S1). Hence, $\mean{v} \approx \Om$ for $|\mean{\cos \beta}| \lesssim 1$ in agreement with the numerical data.  Since $\mean{\cos \beta} < 0$, an increasing maneuverability increases the propulsion velocity according to Eq.~\eqref{eq:langevin_v}, and $\mean{v}$ can exceed $v_0$ by far. However, this does not imply a more efficient pursuit, since the iAOUP velocity is given by $\dot r$ (Eq.~\eqref{eq:langevin_r}), which is then determined by the translation noise. Taylor expansion of the Bessel functions for large arguments ($\Om \gg 1$) and insertion of Eq.~\eqref{eq:mean_v_small} yields
\begin{align} \label{eq:mean_cos_pe_small} 
|\mean{\cos (\beta)}| \approx 1-1/(4 \Om^2) 
\end{align}
for $\Om >1$, and $|\mean{\cos (\beta)}| \approx \sqrt{\pi} \, \Om/2$  for $\Om \to 0$, in agreement with Fig.~\ref{fig:mean_distance}(a). The latter corresponds to  the above uniform distribution of the bearing angle in absence of steering. 

The radial dynamics (Eq.~\eqref{eq:langevin_r}) is predominately diffusive, because thermal noise dominates over the active term. Our analytical calculations yield, with Eq.~\eqref{eq:mean_v_small}, the approximate dependence  (Sec.~S-III.C.)
\begin{equation}
\label{eq:mean_r}
\mean{r} = \frac{2}{\mean{v} \mathrm{Pe} \, |\langle\cos\beta\rangle|}  = \frac{2 \Om}{\mean{v}^2 \mathrm{Pe}}  \, .
\end{equation} 
As displayed in Fig.~S2, this expression describes the simulation data very well over  wide range of maneuverability, when the values of $\mean{v}$ from simulations are inserted.  In the limit $\Om \gg 1$, $\mathrm{Pe} \, \Om <1$, and with $\mean{v} =\Om$, we obtain
\begin{equation} \label{eq:mean_r_lowPe}
\mean{r} = \frac{2}{\mathrm{Pe} \, \Om }  \, ,
\end{equation}   
in agreement with simulation results (Fig.~\ref{fig:mean_distance}). Thus, the radial distance increases with decreasing P\'eclet number and maneuverability. Interestingly, the presence of the maneuverability in Eq.~\eqref{eq:mean_r_lowPe} is in strong contrast to the corresponding behavior of iABPs, where  $\mean{r} $ is independent of $\Omega$ for $\mathrm{Pe}/\sqrt{\Omega} <1$. The dependence of the propulsion velocity qualitatively changes the radial distribution of iAOUPs compared to iABPs \cite{goh:22}. Although, the large propulsion velocity, $\mean{v} \gg1$, implies an effectively larger P\'eclet number, the factor $\mathrm{Pe} \mean{v} \mean{\cos \beta}$ is still much smaller than unity for $\mathrm{Pe} \, \Om \ll 1$, and thermal noise determines the radial distribution function (SM, Figs.~S6, S8).   

\subsection{Large P\'eclet number}

For large $\Pe$, two scaling regimes appear, depending on the ratio of $\Pe/\Om$. In the case  $\mathrm{Pe}  \gg \Om \gg 1$, active propulsion dominates over steering and thermal noise, and the iAOUP behaves as a simple AOUP in the limit $\Pe \to \infty$.  In particular, the propulsion direction is random (Fig.~\ref{fig:mean_distance}(a)), and the bearing angle is determined by active noise, implying a nearly uniform distribution of $\beta$ and $|\mean{\cos \beta}| \ll 1$. Similarly, the equation of motion for $v$ is independent of steering for $\Om |\mean{\cos \beta}|  \ll 1$  and the value $\mean{v} = \sqrt{\pi}/2$ of a simple AOUP is assumed (Fig.~\ref{fig:mean_distance}(b)) (cf.  S-III.C.). Since  $\mean{v} >1$, and with the assumption  $|\mean{\cos(\beta)}| >0$, the radial distance $r$  is determined by propulsion, and noise can be neglected. Then, we obtain the average radial distance (Sec.~S-III.C.)
\begin{equation}
\label{eq:mean_r_high}
\mean{r} = 2 \frac{\mean{v}}{\mean{1/v}} \frac{\mathrm{Pe}}{\Om}  = \frac{\mathrm{Pe}}{\Omega}\, ,
\end{equation}
with $\mean{v}/\mean{1/v}  = 1/2$ (Eq.~(S8)). This power-law is in excellent agreement with the simulation data in  Fig.~\ref{fig:mean_distance}.  Not surprisingly, the iAOUP exhibits the same $\Pe/\Om$ dependence as an iABP \cite{goh:22}, because the random orientation of the propulsion direction for $\mathrm{Pe} \gg \Om \gg 1$ decouples the propulsion velocity from the other variables, and, on average, the iAOUP behaves similar to an iABP, yet with the mean velocity  $\mean{v} = \sqrt{\pi}/2$ and $\mean{v^2} = 1$. Even more, there is a quantitative difference to an iABP, whose mean radial distance is twice as large. Thus, the fluctuations in the magnitude of the propulsion velocity lead to a closer approach of the iAOUP to the target.  The pursuer traverses rosette-like trajectories as displayed in Fig. \ref{fig:sketch_geom}(b). Due to the highly persistent motion ($\Pe/\Omega \sim v_0^2/C \gg 1$), the pursuer overshoots the target, but steering enforces a movement back toward the target.

In the case  $\Om \gg \mathrm{Pe}  \gg 1$, the dynamics of the propulsion velocity and bearing angle are determined predominately by steering rather than by active propulsion. Our analytical calculations yield the approximated average values (Eqs. (S29), (S30))
\begin{align}
\label{eq:cos_universal}
|\mean{\cos\beta}| =   \frac{1}{\sqrt{\mathrm{Pe}\, \Om}} \ , \ \ \ 
\mean{v} = \sqrt{\frac{\Om}{\mathrm{Pe}}} \, .
\end{align}
These power laws are in very good agreement with the simulation results in Figs.~\ref{fig:mean_distance}(a),(b). Interestingly,  thermal noise determines the radial dynamics, because  $ \mean{v}  \Pe \mean{\cos\beta} \approx -1$, and the active term in Eq.~\eqref{eq:langevin_r} is of order unity. Then, the Fokker-Planck equation (S18) for the radial distance  gives 
\begin{align} \label{eq:mean_r_plateau}
\mean{r} \approx 2 \, ,
\end{align} 
in qualitative agreement with simulation results (Fig.~\ref{fig:mean_distance}) --- the actual  value $\mean{r}$ is slightly smaller, because the approximate exponential distribution deviates somewhat from the proper distribution function (Fig.~S6). 
The intersection of the power law \eqref{eq:mean_r_lowPe} with the constant of  Eq.~\eqref{eq:mean_r_plateau} suggest that the plateau $\mean{r} = const.$ extends over the range $1/\Om^2 \ll \mathrm{Pe}/\Om \ll 1$. Thus, there is a broad range of P\'eclet numbers, where a minimum  average pursuer-target distance is assumed, independent of $\mathrm{Pe}$ and $\Om$.   This is in contrast to iABPs, where the smallest average distance is assumed for $\mathrm{Pe}/\sqrt{\Om}=1$ \cite{goh:22}. 

The freedom to adopt the propulsion velocity leads to a novel pursuit dynamics, and can provide an advantage of iAOUPs over iABPs, because the minimal average approach is not very sensitive to the actual values of $\mathrm{Pe}$ and $\Omega$ over a wide range of these values. However, the iAOUP, in average, cannot get arbitrarily close to the target,  whereas the iABP minimal average distance $\mean{r} \sim 1/ \sqrt{\Om}$ decreases with increasing $\Om$.  The intimate coupling of the bearing angle and propulsion velocity allows for an efficient local adaption of propulsion toward the target, where,  at a given $\Pe$ and an increasing $\Om$, the bearing angle is increasingly  randomized  and  the average propulsion velocity increases. This leads to a fast exploration of the neighborhood of the target.  As a result, the coupling implies an effective thermal-like radial dynamics as in a linearly radial-outward growing potential, i.e., an attractive radial force field (SM, Fig.~S8). 

\section{Results: Pursuit by an iAOUP under confinement} 

\begin{figure*}
\includegraphics[width = \textwidth]{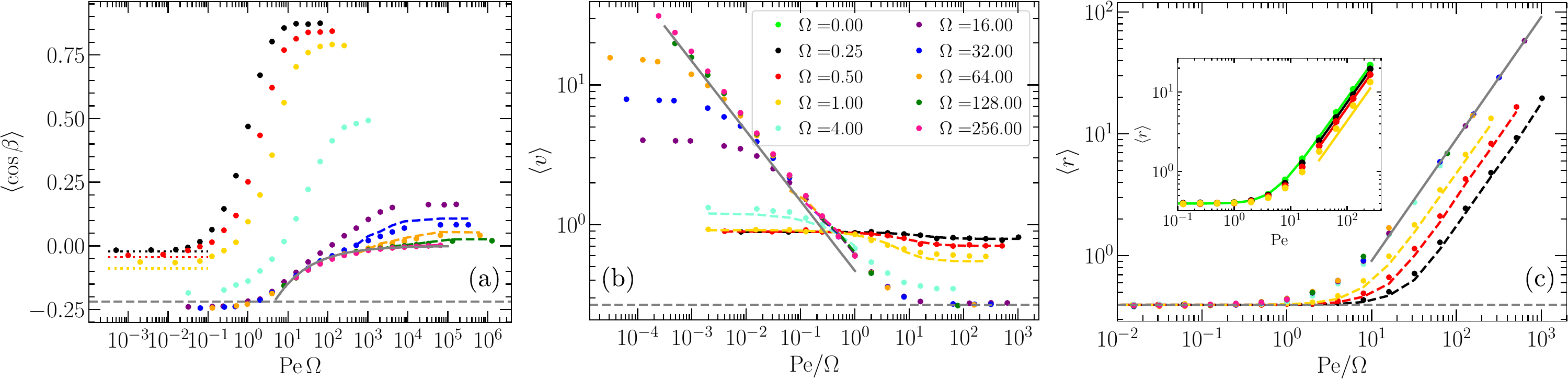}
\caption{Stationary-state properties of  iAOUPs in  harmonic confinement. 
(a) Mean bearing angle $\mean{\cos\beta}$ as a function of $\mathrm{Pe}$ for various $\Om$. The dotted lines for $\mathrm{Pe} \Om \ll 1$ represent Eq.~\eqref{eq:mean_cos_small_om}, and the dashed lines for $\mathrm{Pe} \Om \gg 1$ correspond to Eq.~\eqref{eq:mean_cos_pot_v}  with velocities from simulations.   The gray dashed  and solid lines represent Eq.~\eqref{eq:mean_cos_large_om} and Eq.~\eqref{eq:meancos_pot_large_pe}, respectively.
(b) Mean self-propulsion velocity $\mean{v}$ as a function of $\mathrm{Pe}$.  The dashed  lines represent Eq.~\eqref{eq:v_pot_om} with $\mean{\cos \beta}$ from simulations (cf. (a)).   The gray dashed  and solid lines represent Eq.~\eqref{eq:mean_v_larg_pe_om} and Eq.~\eqref{eq:meanv_pot_large_pe}, respectively.
(c) Mean distance $\mean{r}$ from the target as a function of $\mathrm{Pe}$. The dashed lines  are obtained from  Eq.~(S50a) with $\mean{\cos \beta}$ from the simulations (cf. (a)). The plateau value for $\mathrm{Pe}/\Om <1$ is given by Eq.~\eqref{eq:mean_r_pot_large_pe} as long as $\Om >1$. The gray solid line represents  Eq.~\eqref{eq:mean_r_larg_pe_om}. The inset displays $\mean{r}$ as a function of $\mathrm{Pe}$ for $\Om \leq 1$. The green line represents Eq.~\eqref{eq:mean_r_pot} for a non-steering particle and the solid lines Eq.~\eqref{eq:r_pot_om_exp}. Characteristic  trajectories are presented in Fig.~S9. (See also movies supplM3 --  supplM6.)}
\label{fig:mean_rvcos_potential}
\end{figure*}

The stationary-state properties of simple ABPs and AOUPs confined in a harmonic potential have been studied analytically and by simulations \cite{tail:09,poto:12,roma:12,szam:14,magg:14,solo:15.2,marc:16.3,sand:17,das:18.1}. Most remarkably, the radial distance of ABPs and AOUPs increases linearly with increasing $\mathrm{Pe}$ for $\mathrm{Pe} \gg 1$, and their propulsion direction is preferentially oriented radially outward ($\mean{\cos\beta}  = 1$) \cite{iyer:2023}. However, AOUPs can explore the whole space around the target due to the varying propulsion velocity with the average $\mean{v} = \sqrt{\pi}/2$ independent of activity \cite{das:18.1}.  Figure~\ref{fig:mean_rvcos_potential} (c) (inset) illustrates the dependence of the average radial distance on the P\'eclet number, where  (Sec.~S-IV.)
\begin{align} \label{eq:mean_r_pot}
\mean{r} = \sqrt{\frac{\pi}{2}} \sqrt{\frac{1}{k} + \frac{\mathrm{Pe}^2}{2k(1+k)} } \, ,
\end{align}
with the strength $k$ of the harmonic potential.

Steering of an iAOUP toward the target leads to a competition between confinement and pursuit. Consequently, with increasing maneuverability, the radial averages of an iAOUP in Fig.~\ref{fig:mean_rvcos_potential}(c) deviate substantially from those of a simple AOUP in a harmonic potential (Fig.~\ref{fig:mean_rvcos_potential}(c) (inset)).  

For an iAOUP, the propulsion velocity and the angle $\beta$ depend only implicitly -- via $r$ -- on the strength of the harmonic potential. Thus, even in the presence of the potential, its mean propulsion velocity is given by Eq.~(S10)  in terms of $\Om$ and $\mean{\cos\beta}$, which is confirmed by comparison with simulation results in Fig.~S3. 

\subsection{Small P\'eclet numbers}

\begin{figure*}[t]
\includegraphics[width = \textwidth]{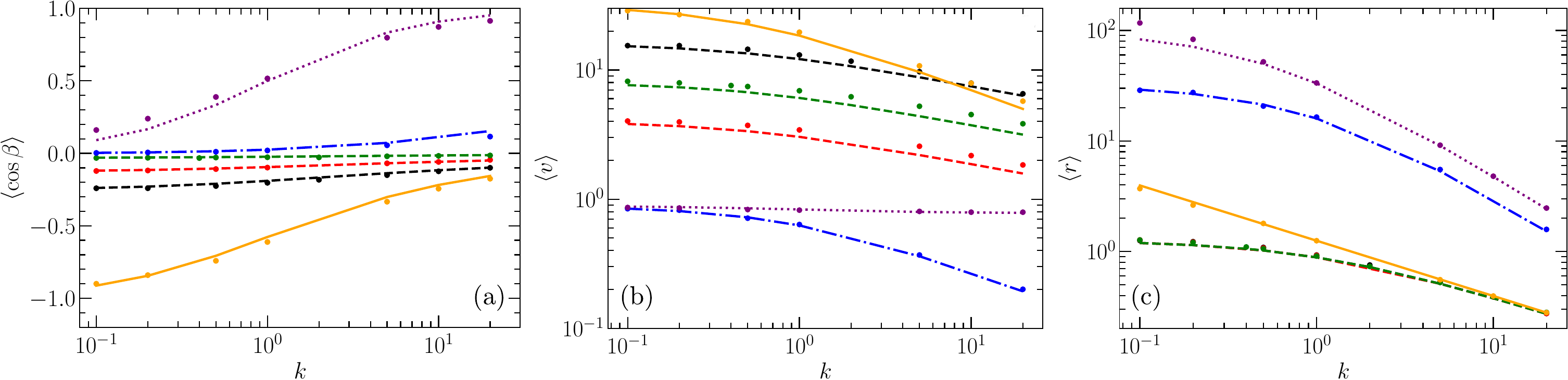}
\caption{Dependencies of the stationary-state averages on the potential strength for the combinations of P\'eclet number and maneuverability:  $(\mathrm{Pe},\Om) =(0.002,32)$ (yellow) (Eqs.~\eqref{eq:mean_r_pot} ($\mathrm{Pe}=0$), \eqref{eq:mean_cos_large_om}, \eqref{eq:mean_v_large_om});   $(\mathrm{Pe},\Om) =(64,0.25)$ (purple) (Eqs.~\eqref{eq:cos_pot_no_om}, \eqref{eq:v_pot_om_k}, \eqref{eq:r_pot_om_exp}); the  
pairs $(\mathrm{Pe},\Om) =(0.25,64), (2,32), (4,256)$  (black, red, green) (Eqs.~\eqref{eq:meancos_pot_large_pe},  \eqref{eq:meanv_pot_large_pe}, \eqref{eq:mean_r_pot_large_pe}); and  $(\mathrm{Pe},\Om) =(1024,32)$ (blue) (Eqs.~\eqref{eq:mean_bvr_larg_pe_om}).
The bullets represent simulation results and the  lines the analytical approximations. 
}
\label{fig:mean_pot_k}
\end{figure*}

In the limit $\mathrm{Pe} \to 0$, the radial distance is decoupled from the propulsion velocity and steering. The dynamics of $r$ is dominated by confinement and thermal noise, and $\mean{r} = \sqrt{\pi/(2k)}$ (Eq.~\eqref{eq:mean_r_pot}). Moreover, the propulsion velocity is independent of steering for $\Omega \ll 1$, with $\mean{v} = \sqrt{\pi}/2$.  Both limits are consistent with the simulation data of Fig.~\ref{fig:mean_rvcos_potential}. As in the absence of confinement, the average $\mean{\cos \beta}$ is given by Eq.~\eqref{eq:mean_cos_low},  but with the argument (Eqs.~(S45), (S46)) 
\begin{align}
z= \frac{\Omega r^2 - v^2 r \mathrm{Pe} }{v+ r^2/(2v)} \, 
\end{align}
at constant $r$ and $v$. For small arguments $z$ ($ 0 < \Omega \ll 1$, $\mathrm{Pe}=0$), Taylor expansion of the Bessel functions yields  
\begin{align}  \label{eq:mean_cos_small_om}
\mean{\cos \beta} \approx  - \frac{\Omega \mean{r}^2}{2 \mean{v}} = -\Om \frac{\sqrt{\pi}}{2k} \, ,
\end{align} 
with the above averages for $\mean{r}$ and $\mean{v}$. Thus, we predict $|\mean{\cos\beta}|$ to decrease linearly with increasing $\Om$, in agreement with Fig.~\ref{fig:mean_rvcos_potential}(a). With increasing  $\Omega$, the tendency of the pursuer to point toward the target increases, and the propulsion velocity becomes $\Omega$ dependent. Insertion of Eq.~\eqref{eq:mean_v_small} into Eq.~\eqref{eq:mean_cos_small_om} yields $\mean{\cos \beta}  = - \sqrt{\pi/(4k)}$ for $\Omega \gg 1$,  a $\mathrm{Pe}$- and $\Om$-independent plateau, qualitatively consistent with the numerical results (Fig.~\ref{fig:mean_rvcos_potential}(a)). By comparison with simulation results (Fig.~\ref{fig:mean_pot_k}(a)), we find that the $k$-dependence is quantitatively more accurately captured by  
\begin{align}  \label{eq:mean_cos_large_om}
\mean{\cos \beta}  = - \frac{1}{\sqrt{2k+1}} \, ,
\end{align}  
which extends the validity of the  average toward small $k$ values.
Hence, despite $\Om \gg 1$,  no perfect alignment toward the target is reached, in contrast to an unconfined iAOUP. 

The mean propulsion velocity  follows from Eqs.~\eqref{eq:mean_v_small}, \eqref{eq:mean_cos_large_om}, and also assumes a $\mathrm{Pe}$-independent  plateau, which is given by 
\begin{align}  \label{eq:mean_v_large_om}
\mean{v} = \frac{\Om}{\sqrt{2k+1}} \, ,
\end{align}
in good agreement with the simulation results. Hence, even the confined pursuer shows an increase of the mean propulsion velocity $\mean{v} \sim \Om$ for $\mathrm{Pe} \to 0$ (Fig.~\ref{fig:mean_rvcos_potential}(b)), yet the average is reduced by the $k$-dependent term. 

Figure~\ref{fig:mean_pot_k} displays the averages $\mean{\cos\beta}$, $\mean{v}$, and $\mean{r}$ as a function of the potential strength $k$. The approximations in Eqs.~\eqref{eq:mean_r_pot} ($\mathrm{Pe}=0$), \eqref{eq:mean_cos_large_om},  and \eqref{eq:mean_v_large_om} capture the $k$-dependence very well.  

In general,  $\mean{\cos \beta}$ increases with increasing $k$ due to confinement, whereas $\mean{v}$ and $\mean{r}$ decrease. Stronger confinement reduces the average orientation of the propulsion direction toward the target, and either randomizes it ($\Om \gg  \Pe$) or enhances the orientation away from the target ($\Pe \gg \Om$). This is particularly pronounced for  $\mathrm{Pe} \gg 1$ and $\Om <1$, where the propulsion orientation changes from a random value to an orientation away for the target, and for $\mathrm{Pe} \ll 1$ and $\Om \gg 1$, where it changes from a strong orientation toward the target to a more random orientation.

\subsection{Large P\'eclet numbers}

The properties of the pursuer at large P\'eclet numbers strongly depend on the maneuverability $\Omega$. For $\Omega < 1$, the averages $\mean{v}$,  $\mean{\cos \beta} >0$, and $\mean{r}$ are close to those of simple AOUPs in the harmonic potential, and decrease with increasing $\Omega$, as displayed in Fig.~\ref{fig:mean_rvcos_potential}. Taylor expansion of the mean propulsion velocity Eq.~(S10)  for $\Omega \mean{\cos \beta} \ll 1$ and $\mean{\cos \beta}  > 0 $ yields (Eq.~(S51))
\begin{align} \label{eq:v_pot_om}
\mean{v} = \frac{\sqrt{\pi}}{2}\left( 1-\frac{4-\pi}{\sqrt{\pi}} \Omega \mean{\cos \beta}\right) .
\end{align}
This expression provides the correct limit  $\mean{v} =\sqrt{\pi}/2$ for $\Omega \to 0$ and the propulsion velocity 
decreases linearly with increasing $\Omega$, in agreement with the numerical results displayed in Fig.~\ref{fig:mean_rvcos_potential}(b).  

In absence of steering, $\Omega =0$,  and thermal noise, we extract the dependence (Eqs.~(S57), (S58))
\begin{align} \label{eq:cos_pot_no_om}
\mean{\cos \beta} =  \mean{\cos \beta}_0 = 1 - \frac{1}{k+1} \, 
\end{align}  
from the Fokker-Planck equation for the radial distance,  in agreement with the simulations results even at small $k$ values. Note that $\mean{\cos \beta}$ is positive, and the propulsion direction points away from the target, which is a consequence of confinement (Fig.~\ref{fig:mean_rvcos_potential}). 
In the limit $\Omega \ll 1$, we expect $\mean{\cos \beta}$ to depended linearly on $\Omega$. Hence, to first order in $\Omega$, we can replace $\mean{\cos\beta}$ in Eq.~\eqref{eq:v_pot_om} by $\mean{\cos\beta}_0$ (Eq.~\eqref{eq:cos_pot_no_om}), which gives 
\begin{align} \label{eq:v_pot_om_k}
\mean{v} = \frac{\sqrt{\pi}}{2}\left( 1-\frac{4-\pi}{\sqrt{\pi}} \Omega \frac{k}{k+1}\right) \, ,
\end{align}
again in good agreement with simulations (Fig.~\ref{fig:mean_rvcos_potential}(b)).
In absence of thermal noise, the radial stationary-state Fokker-Planck equation (S47) yields the mean radial distance (Eq.~(S56))
\begin{align} \label{eq:r_pot_om}
 \mean{r} = \frac{\mean{v}\mean{\cos \beta} }{k} \mathrm{Pe}  .
\end{align}
Figure S4 shows that $\mean{r}$ agrees well with the simulation results, when the averages of the simulation data for $\mean{\cos \beta}$ and $\mean{v}$ are used.  Insertion of the $k$ dependencies of Eq.~\eqref{eq:v_pot_om} and \eqref{eq:cos_pot_no_om} gives 
\begin{align} \label{eq:r_pot_om_exp}
 \mean{r} = \frac{\sqrt{\pi} \, \mathrm{Pe}}{2 k+1} \left( 1 - \frac{4 - \pi}{\sqrt{\pi}} \Omega \frac{k}{k+1}\right)  \, ,
\end{align}
with a correction for small $k$.
The simulation results of Fig.~\ref{fig:mean_rvcos_potential} confirm the analytical predictions of Eq.~\eqref{eq:v_pot_om_k} and \eqref{eq:r_pot_om_exp}. 

As long as $\Omega \lesssim1$, steering only weakly affects the properties of the confined pursuer. Steering toward the target enhances the orientation of  the propulsion direction toward the target, and consequently, implies a decreasing  mean radial distance. The manifestation of the steering correction depends on the strength of the potential, as displayed in Fig.~\ref{fig:mean_pot_k}. Evidently, stronger confinement enhances the radial outward orientation of the propulsion direction. Qualitatively, this behavior is in strong contrast to that in the limit $\mathrm{Pe} \to 0$ and $\Om \gg 1$.

In the limit $\mathrm{Pe}$, $\Omega \gg 1$, and $\mathrm{Pe}/ \Omega <1$ --- here steering dominates over propulsion ---, the active term in Eq.~\eqref{eq:langevin_r} exceeds the contribution by the potential. Hence, as in the absence of confinement,  $\mean{r}$ is approximately given by  Eq.~\eqref{eq:mean_r_high}, in terms of the averages over the velocity, and  (Eq.~(S62))
\begin{align} \label{eq:mean_cos_pot_v}
\mean{\cos \beta} = - \frac{\mean{1/v} \Omega}{(\mean{v} \mathrm{Pe})^2} + \frac{2 k}{\mean{1/v} \Omega}  \, ,
\end{align}   
with an additional contribution by the potential. Applying the approximation $\mean{1/v} \approx 1/\mean{v}$ and inserting Eq.~\eqref{eq:mean_v_small}, we obtain 
\begin{align} \label{eq:meancos_pot_large_pe}
\mean{\cos \beta} =  - \frac{1}{\sqrt{\Omega \mathrm{Pe} }} \frac{1}{\sqrt[4]{1+2k}} \, ,
\end{align}
and by insertion of Eq.~\eqref{eq:meancos_pot_large_pe} into Eq.~\eqref{eq:mean_v_small} 
\begin{align} \label{eq:meanv_pot_large_pe}
\mean{v} = \sqrt{\frac{\Omega}{\mathrm{Pe}}} \frac{1}{\sqrt[4]{1+2k}} \, .
\end{align}
As displayed in Fig.~\ref{fig:mean_rvcos_potential}(a), (b), the numerically obtained dependence is well captured by this expression. In addition,  Eq.~\eqref{eq:mean_r_high} yields $\mean{r} = 2/\sqrt{2k+1}$, a radial distance independent of $\mathrm{Pe}$ and $\Omega$. This relation is in qualitative agreement with the mean radial distance  (Eq.~\eqref{eq:mean_r_pot}) in absence of steering and $\mathrm{Pe}=0$ for $k \gg 1$. Guided by  Eq.~\eqref{eq:mean_r_pot}, quantitative agreement with simulations is obtained for the choice   (Fig.~\ref{fig:mean_rvcos_potential}(c),  \ref{fig:mean_pot_k}(c)) 
\begin{align} \label{eq:mean_r_pot_large_pe}
\mean{r} = \sqrt{\frac{\pi}{2k+1}} \, .
\end{align}
Steering dominates propulsion and the pursuer exhibits the same scaling dependence of the considered averages as a  free iAOUP. All three expressions exhibit  the same $\mathrm{Pe}$ and $\Omega$ dependence as the unconfined pursuer, yet the magnitude is significantly reduced by the respective $k$-dependent term (cf. Fig.~\ref{fig:mean_pot_k}).

In the limit $\mathrm{Pe} \gg \Omega \gg 1$, propulsion dominates over steering and thermal noise. Nevertheless, steering influences the $k$-dependent averages (Fig.~\ref{fig:mean_rvcos_potential}). Qualitatively, this follows from the stationary-state Fokker-Planck equations of the respective variables including the potential  (SM, Sec. S-IV.D.2.). We obtain similar approximate scaling relations as in absence of the potential,
\begin{subequations}
\label{eq:mean_bvr_larg_pe_om} 
\begin{align} \label{eq:mean_b_larg_pe_om}
\mean{\cos \beta} = &  \ \frac{1}{\Om} \frac{2 k}{\sqrt{\pi(k+1)}} \, , \\ \label{eq:mean_v_larg_pe_om}
 \mean{v} = & \ \frac{\sqrt{\pi}}{2} \frac{1}{\sqrt{k+1}} \, , \\ \label{eq:mean_r_larg_pe_om}
 \mean{r}  = & \ \frac{\mathrm{Pe}}{\Omega (k+1)} \, ,
\end{align} 
\end{subequations}
however, with adjusted constants and extensions toward smaller $k$ to match the simulation results (Fig.~\ref{fig:mean_rvcos_potential}). Equation \eqref{eq:mean_v_larg_pe_om} yields the correct asymptotic value for $k \to 0$.  Remarkable is the fact that $\mean{\cos \beta}$ is positive, and that $\Om \mean{\cos \beta}$ (Fig.~S5) and $\mean{v}$ assume a $\mathrm{Pe}$- and $\Om$-independent plateau.   

As displayed in Fig.~\ref{fig:mean_pot_k}, the propulsion direction is random for $k<1$, but points preferentially radially outward with increasing  $k$. In contrast, $\mean{v}$ and $\mean{r}$ decrease gradually with increasing $k$, and $\mean{v}$ is smaller than $\sqrt{\pi}/2$, the value of a simple AOUP in confinement (Fig.~\ref{fig:mean_rvcos_potential}(b), (c)). However,   Eq.~\eqref{eq:mean_r_larg_pe_om} applies for $\mathrm{Pe}/\Om \gg (k+1)/\sqrt{k}$ only. Thus, the crossover from the constant plateau $\mean{r} = \sqrt{\pi/(2k+1)}$ to the universal increase in Eq.~\eqref{eq:mean_r_larg_pe_om}
shifts to larger $\mathrm{Pe}/\Om \sim \sqrt{k}$ ($k \gg 1$) ratios with increasing $k$, i.e., the range of the minimal radial distance extends to large $\mathrm{Pe}/\Om$.

\section{Conclusions}  The properties of iAOUPs differ significantly from those of iABPs. The additional degree of freedom, $v$, of the propulsion velocity is strongly affected by steering and confinement, and in turn influences the radial distance distribution. This leads to qualitatively different scaling relations of unconfined pursuers for their mean radial distance in the regime $\mathrm{Pe}/\Om < 1, \Pe \gg 1$. In contrast, for $\Pe \gg \Om \gg 1$, the same scaling behavior as for iABPs is obtained (Fig.~\ref{fig:mean_distance}). The average radial distance, $\mean{r}$, exhibits a wide $\mathrm{Pe}$- and $\Om$-independent minimum-distance regime for $\Om \gg \Pe \gg 1$, which allows for a close approach of the pursuer over a broad range of activities, i.e., closest approach is rather insensitive to $\mathrm{Pe}$ and $\Om$.  

Confinement strongly affects the propulsion velocity, the bearing angle,  as well as the radial distance, which is reflected in the scaling relations. Specifically for $\Pe \gg \Om \gg 1$, the competition between  propulsion, steering, and confinement leads to an iAOUP behavior rather similar to the unconfined situation, but with scaling relations depending on the strength of confinement, and a broadening of the range of minimal approach. We hope that our theoretical results will contribute to the design of novel self-steering microbots.


\begin{thebibliography}{10}
\expandafter\ifx\csname url\endcsname\relax\def\url#1{\texttt{#1}}\fi

\bibitem{vics:12}
\Name{Vicsek T. \and Zafeiris A.} \REVIEW{Phys. Rep.}{517}{2012}{71}.

\bibitem{jike:15}
\Name{Jikeli J.~F., Alvarez L., Friedrich B.~M., Wilson L.~G., Pascal R., Colin
  R., Pichlo M., Rennhack A., Brenker C. \and Kaupp U.~B.} \REVIEW{Nat.
  Commun.}{6}{2015}{7985}.

\bibitem{bech:16}
\Name{Bechinger C., Di~Leonardo R., L{\"o}wen H., Reichhardt C., Volpe G. \and
  Volpe G.} \REVIEW{Rev. Mod. Phys.}{88}{2016}{045006}.

\bibitem{elge:15}
\Name{Elgeti J., Winkler R.~G. \and Gompper G.} \REVIEW{Rep. Prog.
  Phys.}{78}{2015}{056601}.

\bibitem{lan:16}
\Name{Lan G. \and Tu Y.} \REVIEW{Rep. Prog. Phys.}{79}{2016}{052601}.

\bibitem{scha:18}
\Name{Schauer O., Mostaghaci B., Colin R., H{\"u}rtgen D., Kraus D., Sitti M.
  \and Sourjik V.} \REVIEW{Sci. Rep.}{8}{2018}{9801}.

\bibitem{seng:21}
\Name{SenGupta S., Parent C.~A. \and Bear J.~E.} \REVIEW{Nat. Rev. Mol. Cell
  Biol.}{22}{2021}{529}.

\bibitem{goh:22}
\Name{Goh S., Winkler R.~G. \and Gompper G.} \REVIEW{New Journal of
  Physics}{24}{2022}{093039}.

\bibitem{couz:05}
\Name{Couzin I.~D., Krause J., Franks N.~R. \and Levin S.~A.}
  \REVIEW{Nature}{433}{2005}{513}.

\bibitem{pala:18}
\Name{Palagi S. \and Fischer P.} \REVIEW{Nat. Rev. Mater.}{3}{2018}{113}.

\bibitem{sitt:17}
\Name{Sitti M.} \Book{Mobile Microrobotics} (The MIT Press, Cambridge, MA)
  2017.

\bibitem{kasp:21}
\Name{Kaspar C., Ravoo B.~J., van~der Wiel W.~G., Wegner S.~V. \and Pernice W.
  H.~P.} \REVIEW{Nature}{594}{2021}{345}.

\bibitem{huan:22}
\Name{Huang T.-Y., Gu H. \and Nelson B.~J.} \REVIEW{Annu. Rev. Control Robot.
  Auton. Syst.}{}{2022}{}.

\bibitem{baeu:18}
\Name{B{\"a}uerle T., Fischer A., Speck T. \and Bechinger C.} \REVIEW{Nat.
  Commun.}{9}{2018}{3232}.

\bibitem{lave:19}
\Name{Lavergne F.~A., Wendehenne H., B{\"a}uerle T. \and Bechinger C.}
  \REVIEW{Science}{364}{2019}{70}.

\bibitem{selm:18}
\Name{Selmke M., Khadka U., Bregulla A.~P., Cichos F. \and Yang H.}
  \REVIEW{Phys. Chem. Chem. Phys.}{20}{2018}{10502}.

\bibitem{zhan:21}
\Name{Zhang J., Alert R., Yan J., Wingreen N.~S. \and Granick S.}
  \REVIEW{Nature Physics}{17}{2021}{961}.

\bibitem{alva:21}
\Name{Alvarez L., Fernandez-Rodriguez M.~A., Alegria A., Arrese-Igor S., Zhao
  K., Kr{\"o}ger M. \and Isa L.} \REVIEW{Nat. Commun.}{12}{2021}{4762}.

\bibitem{fodo:16}
\Name{Fodor {\'E}., Nardini C., Cates M.~E., Tailleur J., Visco P. \and van
  Wijland F.} \REVIEW{Phys. Rev. Lett.}{117}{2016}{038103}.

\bibitem{das:18.1}
\Name{Das S., Gompper G. \and Winkler R.~G.} \REVIEW{New J.
  Phys.}{20}{2018}{015001}.

\bibitem{risk:89}
\Name{Risken H.} \Book{The Fokker-Planck Equation} (Springer, Berlin) 1989.

\bibitem{barb:16}
\Name{Barberis L. \and Peruani F.} \REVIEW{Phys. Rev.
  Lett.}{117}{2016}{248001}.

\bibitem{bast:20}
\Name{Bastien R. \and Romanczuk P.} \REVIEW{Sci. Adv.}{6}{2020}{eaay0792}.

\bibitem{negi:22}
\Name{{Singh Negi} R., Winkler R.~G. \and Gompper G.} \REVIEW{Soft
  Matter}{18}{2022}{6167}.

\bibitem{zano:21}
\Name{Zanovello L., Faccioli P., Franosch T. \and Caraglio M.} \REVIEW{J. Chem.
  Phys.}{155}{2021}{084901}.

\bibitem{wyso:22}
\Name{Wysocki A., Dasanna A.~K. \and Rieger H.} \REVIEW{New J.
  Phys.}{24}{2022}{093013}.

\bibitem{eise:17}
\Name{Eisenstecken T., Gompper G. \and Winkler R.~G.} \REVIEW{J. Chem.
  Phys.}{146}{2017}{154903}.

\bibitem{cara:22}
\Name{Caraglio M. \and Franosch T.} \REVIEW{Phys. Rev.
  Lett.}{129}{2022}{158001}.

\bibitem{tail:09}
\Name{Tailleur J. \and Cates M.~E.} \REVIEW{EPL}{86}{2009}{60002}.

\bibitem{poto:12}
\Name{Pototsky A. \and Stark H.} \REVIEW{EPL}{98}{2012}{50004}.

\bibitem{roma:12}
\Name{Romanczuk P., B{\"a}r M., Ebeling W., Lindner B. \and Schimansky-Geier
  L.} \REVIEW{Eur. Phys. J. Spec. Top.}{202}{2012}{1}.

\bibitem{szam:14}
\Name{Szamel G.} \REVIEW{Phys. Rev. E}{90}{2014}{012111}.

\bibitem{magg:14}
\Name{Maggi C., Paoluzzi M., Pellicciotta N., Lepore A., Angelani L. \and
  Di~Leonardo R.} \REVIEW{Phys. Rev. Lett.}{113}{2014}{238303}.

\bibitem{solo:15.2}
\Name{Solon A.~P., Cates M.~E. \and Tailleur J.} \REVIEW{Eur. Phys. J. Spec.
  Top.}{224}{2015}{1231}.

\bibitem{marc:16.3}
\Name{Marconi U. M.~B., Gnan N., Paoluzzi M., Maggi C. \and Di~Leonardo R.}
  \REVIEW{Sci. Rep.}{6}{2016}{23297}.

\bibitem{sand:17}
\Name{{Sandford} C., {Grosberg} A.~Y. \and Joanny J.-F.} \REVIEW{Phys. Rev.
  E}{96}{2017}{052605}.

\bibitem{iyer:2023}
\Name{Iyer P., Winkler R.~G., Fedosov D.~A. \and Gompper G.}
  \REVIEW{arXiv:2209.07880}{}{2022}{}.

\end{thebibliography}

\end{document}